\documentclass[a4paper,USenglish]{article}

\usepackage[T1]{fontenc} %Not needed by LuaLaTeX or XeLaTeX

\usepackage{newunicodechar}
\newunicodechar{≤}{\ensuremath{\leq}}
\newunicodechar{≥}{\ensuremath{\geq}}

\usepackage{booktabs}
\usepackage{multirow}
\usepackage{amsmath}
\usepackage{graphicx}
\usepackage{adjustbox}
\usepackage{hyperref}
\usepackage{cleveref}
\usepackage[backend=biber]{biblatex}
\usepackage{siunitx}
\usepackage[frozencache=true,cachedir=minted-cache]{minted}
\usepackage{csquotes}

\title{The LLM Pro Finance Suite: Multilingual Large Language Models for Financial Applications}
\author{Gaëtan Caillaut, Raheel Qader, Jingshu Liu, Mariam Nakhlé\\Arezki Sadoune, Massinissa Ahmim, Jean-Gabriel Barthelemy \\ Dragon LLM}
\date{\today}

\begin{document}

\maketitle

\begin{abstract}
    The financial industry's growing demand for advanced natural language processing (NLP) capabilities has highlighted the limitations of generalist large language models (LLMs) in handling domain-specific financial tasks. To address this gap, we introduce the LLM Pro Finance Suite, a collection of five instruction-tuned LLMs (ranging from 8B to 70B parameters) specifically designed for financial applications. Our approach focuses on enhancing generalist instruction-tuned models, leveraging their existing strengths in instruction following, reasoning, and toxicity control, while fine-tuning them on a curated, high-quality financial corpus comprising over \SI{50}{\percent} finance-related data in English, French, and German.

    We evaluate the LLM Pro Finance Suite on a comprehensive financial benchmark suite, demonstrating consistent improvement over state-of-the-art baselines in finance-oriented tasks and financial translation. Notably, our models maintain the strong general-domain capabilities of their base models, ensuring reliable performance across non-specialized tasks. This dual proficiency, enhanced financial expertise without compromise on general abilities, makes the LLM Pro Finance Suite an ideal drop-in replacement for existing LLMs in financial workflows, offering improved domain-specific performance while preserving overall versatility. We publicly release two 8B-parameters models to foster future research and development in financial NLP applications: \url{https://huggingface.co/collections/DragonLLM/llm-open-finance}.
\end{abstract}

\section{Introduction}\label{sec:introduction}

The rapid advancement of large language models (LLMs) has revolutionized natural language processing, yet their application in the financial domain remains underexplored. While generalist models like Qwen~\autocite{yang2025qwen3} or Llama~\autocite{grattafiori2024llama} have demonstrated impressive capabilities across diverse tasks, their performance in finance-specific scenarios (such as financial reporting analysis, risk assessment, or regulatory compliance) often falls short due to the unique linguistic and domain-specific nuances of financial text.

The financial industry has increasing needs of large language models to process complex and domain-specific text from regulatory filings to market analysis. However, the development of specialized financial LLMs remains fragmented, with most efforts focused on proprietary systems or narrow applications. Notable attempts include BloombergGPT~\autocite{wu2023bloomberggpt}, a proprietary model fine-tuned on Bloomberg’s financial datasets, and FinMA which has been trained as part of the PIXIU~\autocite{xie2023pixiu} project. Additionally, other models such as FinBERT~\autocite{yang2020finbert,araci2019finbert,liu2021finbert,huang2023finbert} and FinGPT~\autocite{yang2023fingpt} have also been developed to address the challenges of financial language processing. Despite these efforts, financial LLMs face significant challenges: limited open access, lack of standardized benchmarks, mostly English-centric, and insufficient evaluation of instruction-tuned or chat-oriented models, which are critical for interactive applications like financial advisory or real-time market analysis. Moreover, as foundation models evolve rapidly, earlier specialized systems risk obsolescence, underscoring the need for adaptable and continuously updated financial LLMs.

As a consequence, our goals are to, first, develop an efficient data curation pipeline, and, second, create LLM that have strong financial knowledge, without sacrificing performances in general tasks. We also aim at preserving their multilingual capabilities by focusing on more languages such as French and German alongside English. Contrary to previous work, which mainly concentrate on finetuning pretrained foundation models, we chose to focus on improving generalist instruction-tuned models. Indeed, instruct, or chat, models have already learned valuable capabilities such as instruction following, reasoning, agentic behavior, and toxicity guards. By building upon these existing strengths, we aim to create more robust and versatile financial LLMs that can handle a wide range of tasks and scenarios, while focusing exclusively on our target: adapting the model to the financial domain. To this aim, we first curate a diverse dataset comprising financial documents and chat-style interactions in English, French and German; ensuring coverage of key subdomains such as corporate finance, investment analysis, and economic policy.
We identified the best candidates through extensive testing, and we trained five models spanning parameter sizes from 8B to 70B:
\begin{itemize}
    \item \textbf{Llama Open Finance 8B} Small sized model without reasoning support, based on \texttt{Llama 3.1 8B Instruct}
    \item \textbf{Qwen Open Finance R 8B} Small sized model with reasoning support, based on \texttt{Qwen 3 8B}
    \item \textbf{Gemma Pro Finance 12B} Small sized model, particularly good at translation related tasks, based on \texttt{Gemma 3 12B it}
    \item \textbf{Qwen Pro Finance R 32B} Medium sized model with reasoning support, based on \texttt{Qwen 3 32B}
    \item \textbf{Llama Pro Finance 70B} Leading performer in most of the tasks, without reasoning support, based on \texttt{Llama 3.1 70B Instruct}
\end{itemize}

Beyond the release of these models, our work contributes to the field of financial large language models in three key ways.
First, we describe the construction of our comprehensive financial dataset that spans diverse tasks and text types, from regulatory disclosures and financial news to conversational data, ensuring broad domain coverage.
Second, we design a financial benchmark suite that integrates both public and in-house evaluation sets, allowing for consistent and transparent assessment across models and tasks.
Finally, we fine-tune a series of instruction-tuned financial LLMs ranging from 8B to 70B parameters, and we show that they are not only better in financial task, but they also preserve their general domain capabilities. We release to the open-source community two 8B models to provide a foundation for future research in financial applications\footnote{\url{https://huggingface.co/collections/DragonLLM/llm-open-finance}}.

\section{Background and Related Work}

\paragraph{Financial Language Models}

The current trend in generative AI is to pretrain large language models (LLMs) on vast and diverse corpora, enabling them to understand a wide range of domains. In pursuit of Artificial General Intelligence (AGI), these models are continuously scaled in both data and parameters. While they now surpass human performance on many common Natural Language Understanding (NLU) benchmarks, their performance in highly specialized fields—such as healthcare, legal, regulatory, or finance—remains limited. Nevertheless, pretrained LLMs serve as powerful foundations as their broad knowledge can be efficiently adapted to specific domains via finetuning.

Specializing language models for financial applications has been a common practice since the emergence of large general-purpose models. For example, numerous practitioners and researchers have proposed finetuned variants of the \texttt{BERT}~\autocite{devlin2019bert}, leading to a range of \texttt{FinBERT} models~\autocite{yang2020finbert,araci2019finbert,liu2021finbert,huang2023finbert} based on an encoder architecture. These models target specific financial tasks that the base model could not adequately address, such as sentiment analysis or information extraction from financial documents.

The rise of decoder-based architectures has, similarly, driven a new wave of financially oriented LLMs. This trend is supported by recent research on scaling laws of decoder-only models~\autocite{caillaut2024scaling} which have demonstrated the potential of these architectures in financial translation tasks. The AI4Finance initiative introduced a framework for finetuning open-source pretrained LLMs and released several financial variants~\autocite{yang2023fingpt}. \textcite{inserte2024large} shows that a carefully finetuned model can outperforms significantly larger models on a specific task. \textcite{xie2023pixiu} proposed a comparable approach under the PIXIU framework, focusing on finetuning the \texttt{Llama} model~\autocite{touvron2023llama}. One notable exception is \texttt{BloombergGPT}~\autocite{wu2023bloomberggpt}, which was trained entirely from scratch on a 700B-token financial dataset.

These efforts collectively highlight that general-purpose pretrained LLMs require adaptation to handle the complexity of financial language for real-world deployment. However, the pace of progress in this field is so fast that earlier specialized models often become obsolete shortly after the release of newer LLM generations. Modern state-of-the-art models are now trained on tens of trillions of tokens using sophisticated training strategies, making them increasingly competent even in niche domains. Yet, as we demonstrate in this paper, room for improvement remains, and the adaptation challenge may grow harder as base models become stronger.

\paragraph{Financial Datasets and Benchmarks}

The performance of AI models is typically assessed using standardized benchmarks. General-purpose benchmarks such as HellaSwag~\autocite{zellers2019hellaswag} and MMLU~\autocite{wang2024mmlu} evaluate broad language understanding, while others like GSM8K~\autocite{cobbe2021gsm8k} focus on mathematics and reasoning ability. Although many well-established benchmarks exist for mainstream tasks, comparatively, few target the financial domain.

FinBen~\autocite{xie2024finben}, introduced under the PIXIU framework~\autocite{xie2023pixiu}, was one of the first large-scale financial benchmarks. It comprises 35 test datasets spanning diverse NLP tasks on financial text.
FinQA~\autocite{chen2021finqa} and ConvFinQA~\autocite{chen2022convfinqa} focus on evaluating numerical reasoning over financial data. These datasets consist of question-answer pairs that often require domain knowledge of financial formulas to solve.
FinEval~\autocite{zhang2023fineval} is another large-scale financial benchmark collection, though it is restricted to Chinese-language tasks.

Other benchmarks focus specifically on retrieval-augmented generation (RAG), such as FinanceBench~\autocite{islam2023financebench}. This dataset is constructed from SEC filings, with associated questions, answers, and supporting evidence. Such benchmarks are increasingly valuable as RAG-based systems gain traction. However, evaluating models in controlled conditions remains challenging, as it often requires an LLM-as-a-Judge setup. This approach introduces ambiguity in judgment criteria, as LLMs may prioritize aspects that are not aligned with the desired outcomes. Furthermore, LLMs may lack the domain-specific expertise required to accurately assess financial tasks, and risk reinforcing existing biases in the training data, potentially resulting in flawed or misleading evaluations.

\section{Data Collection}\label{sec:data-collection}

The LLM Pro Finance models are trained on a high-quality and diverse dataset specifically designed to cover the domains of finance, economics, and business. As shown in \Cref{tab:data-distribution}, the dataset has been build in order to cover a large spectrum of financial tasks while remaining diverse enough to maintain strong multilingual capabilities and reduce the risk of catastrophic forgetting~\autocite{ramasesh2021effect,zhai2023investigating,luo2025empirical,aleixo2023catastrophic}. The dataset can be divided in fives categories:

\begin{description}
    \item[Finance] Over half (\SI{54.4}{\percent}) of the dataset consists of financial content, including market analysis, regulatory documents, accounting data, and synthetic financial discussions. This ensures strong domain expertise in banking, investment, and economic reasoning.
    \item[Translation] Nearly \SI{20}{\percent} of the data is dedicated to translation tasks, supporting robust multilingual capabilities in both general and financial domains.
    \item[General] \SI{15.6}{\percent} covers broad topics (e.g., Wikipedia, scholarly articles) to maintain general-world knowledge and safety alignment.
    \item[RAG] \SI{8}{\percent} is allocated to retrieval-augmented generation (RAG), enhancing factual accuracy and reducing hallucinations.
    \item[Math, Code and Agentic behavior] \SI{2.2}{\percent} focuses on logical reasoning, math, code and external tool-calling for technical problem-solving.
\end{description}

This balanced composition enables the model to excel in financial applications while retaining versatility for general and multilingual use cases. As demonstrated by \textcite{ke2025demystifying}, adapting large language models (LLMs) to domain-specific tasks is better done by trainig the model on an hybrid dataset combining Continuous Pre-Training (CPT) data (raw text documents) and Supervised Fine-Tuning (SFT) data (instruction-following or conversational data). CPT data enables the model to absorb domain-specific concepts, while SFT data preserves critical capabilities like instruction following, reasoning, and toxicity mitigation. To balance these aspects, we collected both types of data while ensuring the model retains general knowledge by including a sufficient amounts of general-domain content. Multilingual capabilities were maintained through additional translation data, even though our primary focus is French, English, and German.

\begin{table}
    \centering
    \begin{tabular}{lrr}
        \toprule
        Category                             & Sample Count  & Ratio               \\
        \midrule
        Finance                              & \num{1102496} & \SI{54.4}{\percent} \\
        Translation                          & \num{402115}  & \SI{19.8}{\percent} \\
        General                              & \num{315981}  & \SI{15.6}{\percent} \\
        RAG (Retrieval-Augmented Generation) & \num{162383}  & \SI{8.0}{\percent}  \\
        Reasoning, Math and Code             & \num{43837}   & \SI{2.2}{\percent}  \\
        \midrule
        Total                                & \num{2026812} & \SI{100}{\percent}  \\
        \bottomrule
    \end{tabular}
    \caption{Composition of LLM Pro Finance's training dataset.}\label{tab:data-distribution}
\end{table}

\subsection{CPT Dataset}

The CPT dataset was constructed through web crawling and automated filtering. Financial data was sourced from websites dedicated to financial, business or economic topics. For instance, we include a substantial portion of the AGEFI\footnote{\url{https://www.agefi.fr/}} articles, ensuring document of high quality and a broad coverage of financial topics.

We also leveraged existing crawling efforts such as \texttt{fineweb}~\autocite{penedo2024the} and \texttt{fineweb-2}~\autocite{penedo2025fineweb2pipelinescale} in order not to cause unnecessary pressure on remote servers and also because these two corpora have already undergone through extensive filtering and cleaning processes.

However, these datasets are mostly suitable for the training of generalist LLM. In order to keep only the financial content, we developed a custom classifier using a two-step process inspired from the \texttt{fineweb-edu}~\autocite{lozhkov2024fineweb-edu} initiative. First, a strong pre-trained LLM (\texttt{Qwen3 235B}) was prompted to label a representative sample of documents from the \texttt{fineweb}~\autocite{penedo2024the} and \texttt{fineweb-2}~\autocite{penedo2025fineweb2pipelinescale} corpora. Then, we finetuned a smaller classifier based on \texttt{Multilingual DeBERTa}~\autocite{he2021debertav3,he2021deberta} to replicate the classification behavior of the strong LLM. Finally, we used this small model (less than 300 millions parameters) to filter financial documents from our large scale web corpora. Additionally, financial articles from Wikipedia were extracted using both the Wikipedia categories (e.g., \enquote{Corporate finance}) and our classifier. The target categories were mined by navigating the Wikipedia categories graph, starting different walks from seed categories. We realized that this approach yields many categories unrelated to finance, so it is not suitable for financial articles extraction. Nevertheless, we decided to keep this \textit{noise} in order to include more general data in the training, as a way to preserve the general knowledge of the base model.

At this stage, we have a large finance focused dataset that should reach a sufficient level of quality to be suitable for \textit{pretraining} an LLM. However, as we aim at finetuning high performing instruction-tuned LLMs, we need an even higher degree of quality. Hence, we filtered the resulting dataset using a strong LLM (\texttt{Qwen3 235B}) with the following predefined set of rubrics~\autocite{pathak2025rubric}:
\begin{minted}[breaklines,tabsize=2,escapeinside=||]{markdown}
|\textbf{Accuracy and Reliability:}| Is the content factually correct and free from misinformation?
|\textbf{Educational Value:}| Is the content structured to help the reader learn or understand something clearly?
|\textbf{Clarity and Structure:}| Is the writing well-organized, logical, and easy to follow?
|\textbf{Writing Quality and Tone:}| Is the tone professional, neutral, and appropriate for educational material?
|\textbf{Originality and Insight:}| Does the document provide thoughtful analysis or useful insight?
\end{minted}
We also asked the LLM Judge to provide an overall rating and to decide if the content needs to be retained or rejected.

\subsection{SFT Dataset}

The SFT dataset combines open-source resources from Hugging Face Hub with synthetic data generated from text documents. To create synthetic instruction-following examples, we deployed multiple strong LLMs to extract key concepts (e.g., financial metrics, regulatory terms) from documents. These concepts were used as seeds to generate multi-turn question-answer (Q/A) pairs in multiple languages (French, English or German). For example, a document about \enquote{Earnings per Share (EPS)} might generate questions like:

\begin{minted}[breaklines,tabsize=2]{text}
    Q: What is the EPS formula?
    A: EPS = Net Income / Number of Shares Outstanding.
\end{minted}

To clean it further, we filtered the resulting, aggregated, dataset using a strong a LLM (\texttt{Qwen3 235B}) and the following predefined set of rubrics~\autocite{pathak2025rubric}:
\begin{minted}[breaklines,tabsize=2,escapeinside=||]{markdown}
|\textbf{Relevance:}| Does the answer directly address the question asked?
|\textbf{Accuracy:}| Is the information in the answer factually correct?
|\textbf{Completeness:}| Does the answer sufficiently cover the important aspects of the question?
|\textbf{Clarity:}| Is the answer clearly worded and easy to understand?
|\textbf{Conciseness:}| Is the answer free of unnecessary information or repetition?
|\textbf{Tone and Appropriateness:}| Is the tone suitable for the context (professional, neutral, etc.)?
|\textbf{Context Alignment:}| Does the style, tone, and length of the answer match the question's intent and complexity?
|\textbf{Instruction Following:}| Does the answer rigorously follow the instruction?
\end{minted}
Similarly to the CPT data filtering process, we also asked the LLM Judge to provide an overall score from 1 (very poor) to 5 (excellent), and we kept only the samples with a score of at least 4.

\subsection{Translation Dataset}

As we aim to maintain robust multilingual capabilities, we put lots of efforts into the translation dataset. We build two kinds of translation datasets: a CPT and an SFT one. The SFT dataset is build using a large collection of in-house parallel data and a set of simple prompts.

The CPT dataset is a \textit{multi-directional translation dataset}. Traditional translation training often focuses on single-language pairs (e.g., English → French), requiring repeated exposure to the same content in both directions. We curated a dataset from the OPUS repository~\autocite{4992de1b5fb34f3e9691772606b36edf} which includes translations across multiple languages (e.g., English, French, German) for the same source text. While not all corpus available in OPUS support this, it is often possible to group all available translation of a single sentences using a combination the bilingual and monolingual files available on OPUS. The result is a CPT translation dataset of where each sample resembles:
\begin{minted}[breaklines,tabsize=2]{text}
    English: English sentence
    French: Phrase en français
    German: Deutscher Satz
\end{minted}

The languages order must be shuffled in order to properly let the model learn multi-directional translation, otherwise the model couldn't learn to translate to English if all samples follow the same languages order as above. This multi-directional approach has the benefit of reducing redundancy while enabling the model to learn multi-directional translations (English → French, French → German, etc.) from a single sample.

Finally, the resulting dataset was filtered using strong a LLM (\texttt{Qwen3 235B}) and the following predefined set of rubrics~\autocite{pathak2025rubric}:
\begin{minted}[breaklines,tabsize=2,escapeinside=||]{markdown}
|\textbf{Accuracy}| How precisely the translation conveys the original meaning. All key information must be preserved
|\textbf{Style and Tone}| The translation should reflect the same tone (formal, informal, technical, poetic, etc.) and style (concise, elaborate, emotional, etc.) as the source text
|\textbf{Casing Consistency}| The translation must follow the same casing conventions as the source text (e.g., sentence case, title case, all caps)
|\textbf{Cultural Appropriateness}| The translation must account for cultural differences and nuances in the target language, avoiding literal renderings that may be unnatural, inappropriate, or confusing.
\end{minted}
Similarly to the CPT and SFT data filtering process, we also asked the LLM Judge to provide an overall score from 1 (very poor) to 5 (excellent), and we kept only the samples with a score of at least 4.

\section{Training and experiments}

The LLM Pro Finance models were finetuned using \texttt{DeepSpeed} with ZeRO stage 3~\autocite{rajbhandari2020zero} and CPU offloading and \texttt{TRL}~\autocite{vonwerra2022trl} on Nvidia H100 GPUs. We aim to cover the wider range of application possible, hence we finetuned a collection of five models spanning size of 8B to 70B parameters. Smaller models are intended to be used in tasks requiring lower latency or high frequency processes, while larger models are better at reasoning, agent workflows or multi-turn conversations.
\begin{itemize}
    \item \textbf{Llama Open Finance 8B} Small sized model without reasoning support, based on \texttt{Llama 3.1 8B Instruct}
    \item \textbf{Qwen Open Finance R 8B} Small sized model with reasoning support, based on \texttt{Qwen 3 8B}
    \item \textbf{Gemma Pro Finance 12B} Small sized model, particularly good at translation related tasks, based on \texttt{Gemma 3 12B it}
    \item \textbf{Qwen Pro Finance R 32B} Medium sized model with reasoning support, based on \texttt{Qwen 3 32B}
    \item \textbf{Llama Pro Finance 70B} Leading performer in most of the tasks, without reasoning support, based on \texttt{Llama 3.1 70B Instruct}
\end{itemize}
For now, we open source only the two 8B models as part of our HuggingFace collection \textbf{LLM Open Finance}\footnote{\url{https://huggingface.co/collections/DragonLLM/llm-open-finance}}.

The models were trained on the traditional causal language modeling loss (i.e. next token prediction task), but we ignored the loss of the user prompt tokens in the SFT data subset. We also ignored the loss of the \SI{15}{\percent} first tokens in the CPT subset. The reason we did that is that we believe it does not make sense to train the model to \textit{start a document from scratch} as, in real world scenarios, the models will always be provided an initial context. Furthermore, as the model is trained on a mix of CPT and SFT data, training the models on the first tokens might confuse the model as it needs to \textit{choose} between generating raw text or chat structured content. Ignoring the first tokens allows us to remove and ignore this uncertainty. However, preliminary experiments showed that ignoring the first tokens has very little effect on the overall performances of the model, but we still decided to ignore them to remain consistent with the SFT data, which always begin with a masked prompt.

All models were trained with a batch size per GPU of \num{1} and gradient accumulation steps of \num{8}. We used the AdamW optimizer with a maximum learning rate set to \num{1e-5}. More details regarding the hardware resources consumed for each model can be found on \Cref{tab:compute-resources}.

\begin{table}[h]
    \centering
    \begin{tabular}{lrrrr}
        \toprule
        Model                  & Params & GPUs & Real Time      & GPU hours         \\
        \midrule
        Llama Open Finance 8B  & 8B     & 32   & \SI{20}{\hour} & \SI{640}{\hour}   \\
        Qwen Open Finance R 8B & 8B     & 32   & \SI{20}{\hour} & \SI{640}{\hour}   \\
        Gemma Pro Finance 12B  & 12B    & 64   & \SI{60}{\hour} & \SI{3840}{\hour}  \\
        Qwen Pro Finance R 32B & 32B    & 128  & \SI{70}{\hour} & \SI{9000}{\hour}  \\
        Llama Pro Finance 70B  & 70B    & 256  & \SI{40}{\hour} & \SI{10000}{\hour} \\
        \bottomrule
    \end{tabular}
    \caption{Compute resources used for each model. Reported times (real and GPU) are approximate.}\label{tab:compute-resources}
\end{table}

\section{Evaluation}

To assess the performance of LLM Pro Finance models, we conducted a comprehensive evaluation using both general domain benchmarks and specialized financial benchmarks. Specifically, we selected FinBen benchmarks and MMLU finance from FinDAP~\autocite{ke2025demystifying}, which are widely recognized for evaluating language models in financial contexts. As these benchmarks are English-only, we translated them to French as we expect the models to be proficiency in both languages. Additionally, we designed an experiment leveraging the LLM-as-a-Judge approach to evaluate the Retrieval-Augmented Generation (RAG) capabilities of the models.

All the benchmarks mentioned in the following sections have been reformatted to match the chat template of the evaluated LLM. We provide a detailed example of the reformatting process in \Cref{app:reformat-to-chat}. We found that it is very important to evaluate the models in the same context as in real-world scenario. As LLM Pro Finance is aimed to be used in a conversational way, it is important for us to evaluate it like this, not like a foundation model. While we generally observe only little discrepancies between these two modes, sometimes the difference is huge. For instance, it is the case on the Human Eval benchmark where the scores in \enquote{foundation mode} are low and do not reflect the real capabilities of the models:
\begin{itemize}
    \item Qwen3 32B scores 0.37 in \enquote{foundation mode} and 0.88 in \enquote{chat mode}
    \item Llama3.1 70B scores 0.66 in \enquote{foundation mode} and 0.78 in \enquote{chat mode}
\end{itemize}

\subsection{Evaluation of general linguistic understanding}

While financial understanding is the core priority of the LLM Pro Finance Suite, it remains important to keep strong general language understanding capabilities. Therefore, we evaluated the models on a set of traditional LLM benchmarks covering general language understanding, general world knowledge and code. As shown in \Cref{tab:general-benchmark}, LLM Pro Finance models generally yield results on-par with their respective baseline, except for HellaSwag on the small variant (LLM Pro Finance almost doubles the score of the base model) and for Human Eval, which is a Coding benchmark. As we did not focus on coding during the training, this result was expected. Nevertheless, our largest model managed to outperform its base model.

This first set of results indicate that LLM Pro Finance models did not lose general knowledge through their financial adaptation.

\begin{table}[ht]
    \centering
    \begin{tabular}{lrrrrrrr}
        \toprule
                   & \multicolumn{2}{c}{Gemma 12B} & \multicolumn{2}{c}{Qwen 32B} & \multicolumn{2}{c}{Llama 70B}                         \\
        \cmidrule(lr){2-3}
        \cmidrule(lr){4-5}
        \cmidrule(lr){6-7}
                   & Pro Fin                       & Base                         & Pro Fin                       & Base & Pro Fin & Base \\
        \midrule
        Hellaswag  & 0.60                          & 0.34                         & 0.61                          & 0.64 & 0.66    & 0.65 \\
        MMLU Pro   & 0.47                          & 0.49                         & 0.66                          & 0.70 & 0.60    & 0.62 \\
        PIQA       & 0.81                          & 0.78                         & 0.80                          & 0.78 & 0.83    & 0.84 \\
        TriviaQA   & 0.77                          & 0.77                         & 0.77                          & 0.75 & 0.89    & 0.89 \\
        Human Eval & 0.78                          & 0.86                         & 0.50                          & 0.88 & 0.80    & 0.78 \\
        \bottomrule
    \end{tabular}
    \caption{Performance of LLM Pro Finance models on non-financial benchmarks. The LLM Pro Finance models are generally on par with their baseline models, indicating that general domain knowledge and natural language understanding capabilities are not degraded through the adaptation to the financial domain.}\label{tab:general-benchmark}
\end{table}

\subsection{Evaluation of financial skills and knowledge}

\paragraph{Financial Acronyms Recognition}{
    We developed an in-house test dataset focused exclusively on financial acronyms in English and French. Financial acronyms are ubiquitous in financial documents and their understanding is critical to accurately interpret user queries. This dataset was designed to quantify the models' knowledge in this specific aspect of the financial sector. While the dataset's principle is straightforward, it effectively serves its purpose by providing a targeted assessment of the models' understanding of financial terminology. A sample looks like this:
    \begin{verbatim}
    What does the acronym "{{acronym}}" stand for?
\end{verbatim}
    And the model answer is expected to contain the expanded acronym.

    \begin{table}[ht]
        \centering
        \begin{tabular}{llrrr}
            \toprule
                                   &      & \multicolumn{2}{c}{Accuracy} &                             \\
            \cmidrule{3-4}
            Model                  & Lang & Pro Fin                      & Base & Improvement          \\
            \midrule
            Llama Open Finance 8B  & EN   & 0.48                         & 0.38 & \SI{28.31}{\percent} \\
            Qwen Open Finance R 8B & EN   & 0.49                         & 0.39 & \SI{23.70}{\percent} \\
            Gemma Pro Finance 12B  & EN   & 0.52                         & 0.42 & \SI{23.91}{\percent} \\
            Qwen Pro Finance R 32B & EN   & 0.51                         & 0.42 & \SI{19.89}{\percent} \\
            Llama Pro Finance 70B  & EN   & 0.52                         & 0.47 & \SI{10.68}{\percent} \\

            \addlinespace

            Llama Open Finance 8B  & FR   & 0.28                         & 0.17 & \SI{64.77}{\percent} \\
            Qwen Open Finance R 8B & FR   & 0.28                         & 0.17 & \SI{61.36}{\percent} \\
            Gemma Pro Finance 12B  & FR   & 0.33                         & 0.23 & \SI{42.74}{\percent} \\
            Qwen Pro Finance R 32B & FR   & 0.34                         & 0.26 & \SI{31.30}{\percent} \\
            Llama Pro Finance 70B  & FR   & 0.37                         & 0.29 & \SI{29.93}{\percent} \\
            \bottomrule
        \end{tabular}
        \caption{Performance of LLM Pro Finance models in understanding financial acronyms, measured by accuracy. The Pro Finance models consistently outperform their baseline in both English and French.}\label{tab:acronym-benchmark}
    \end{table}

    Performance of LLM Pro Finance models on this financial acronym understanding benchmark are shown in \Cref{tab:acronym-benchmark}.
    It clearly shows that all the LLM Pro Finance models improve over the baseline by a large margin, especially in French. This is clearly because of the vast amount of English data used to train the base model, and prove that work still need to be carried out to reach English-level performances on all languages. The accuracy on the English acronyms seem to stagnate around \num{0.50} for all finetuned models, with smaller models outperforming larger base models.
}

\paragraph{Extensive Financial Benchmarking}{
    \begin{table}[ht]
        \centering
        \begin{tabular}{lrrrrrr}
            \toprule
                                        & \multicolumn{2}{c}{Llama 70B} & \multicolumn{2}{c}{Qwen 32B} & \multicolumn{2}{c}{Gemma 12B}                                                          \\
            \cmidrule(lr){2-3}
            \cmidrule(lr){4-5}
            \cmidrule(lr){6-7}
            Benchmark                   & Pro Fin                       & Base                         & Pro Fin                       & Base             & Pro Fin          & Base             \\
            \midrule
            MMLU Finance                & \underline{0.83}              & 0.81                         & \textbf{0.86}                 & 0.80             & 0.75             & 0.75             \\
            Acronyms                    & \textbf{0.52}                 & 0.47                         & \underline{0.51}              & 0.42             & \textbf{0.52}    & 0.42             \\
            FinBen finarg arc           & 0.56                          & \underline{0.63}             & \textbf{0.65}                 & \underline{0.63} & 0.53             & 0.54             \\
            FinBen finarg auc           & 0.63                          & \textbf{0.68}                & 0.63                          & 0.55             & 0.65             & \underline{0.67} \\
            FinBen fiqasa               & \textbf{0.86}                 & 0.68                         & \underline{0.85}              & 0.44             & 0.83             & 0.69             \\
            FinBen fomc                 & 0.63                          & \textbf{0.68}                & 0.65                          & 0.66             & \underline{0.67} & \underline{0.67} \\
            FinBen fpb                  & 0.77                          & \underline{0.81}             & \textbf{0.84}                 & 0.80             & 0.68             & 0.78             \\
            FinBen headlines            & 0.78                          & 0.77                         & 0.77                          & 0.73             & \underline{0.79} & \textbf{0.81}    \\
            FinBen ma                   & \textbf{0.86}                 & 0.83                         & \underline{0.85}              & 0.83             & 0.84             & 0.79             \\
            FinBen mlesg                & 0.45                          & \textbf{0.48}                & 0.45                          & 0.41             & \underline{0.46} & 0.44             \\
            FinBen multifin             & \textbf{0.76}                 & \underline{0.72}             & 0.71                          & 0.65             & 0.71             & 0.67             \\
            FinBen ner                  & \underline{0.27}              & 0.16                         & \textbf{0.30}                 & 0.19             & 0.12             & 0.12             \\
            FinBen tsa \( \downarrow \) & 0.28                          & 0.26                         & 0.25                          & \textbf{0.22}    & \underline{0.23} & \underline{0.23} \\
            \midrule
            Ranked first                & \textbf{4}                    & \underline{3}                & \textbf{4}                    & 1                & 1                & 1                \\
            \bottomrule
        \end{tabular}
        \caption{English financial tasks. Best and second scores are, respectively, bold and underlined. The last row \enquote{Ranked first} show the number of time the model is ranked first.}\label{tab:en-finance-benchmarks}
    \end{table}

    \begin{table}[ht]
        \centering
        \begin{tabular}{lrrrrrr}
            \toprule
                                        & \multicolumn{2}{c}{Llama 70B} & \multicolumn{2}{c}{Qwen 32B} & \multicolumn{2}{c}{Gemma 12B}                                                          \\
            \cmidrule(lr){2-3}
            \cmidrule(lr){4-5}
            \cmidrule(lr){6-7}
            Benchmark                   & Pro Fin                       & Base                         & Pro Fin                       & Base             & Pro Fin          & Base             \\
            \midrule
            MMLU Finance                & 0.77                          & \underline{0.78}             & \textbf{0.82}                 & 0.75             & 0.70             & 0.70             \\
            Acronyms                    & \textbf{0.37}                 & 0.29                         & \underline{0.34}              & 0.26             & 0.33             & 0.23             \\
            FinBen finarg arc           & 0.64                          & \underline{0.65}             & \underline{0.65}              & \textbf{0.66}    & 0.47             & 0.64             \\
            FinBen finarg auc           & \underline{0.69}              & 0.62                         & \textbf{0.70}                 & 0.67             & 0.68             & 0.67             \\
            FinBen fiqasa               & \textbf{0.78}                 & 0.69                         & \underline{0.77}              & 0.57             & \textbf{0.78}    & 0.71             \\
            FinBen fomc                 & 0.56                          & \underline{0.58}             & 0.56                          & \textbf{0.60}    & \underline{0.58} & 0.57             \\
            FinBen fpb                  & \textbf{0.82}                 & 0.77                         & \underline{0.81}              & 0.75             & 0.76             & 0.79             \\
            FinBen headlines            & \underline{0.80}              & \textbf{0.81}                & 0.76                          & 0.73             & 0.75             & \underline{0.80} \\
            FinBen ma                   & \textbf{0.86}                 & \underline{0.85}             & \underline{0.85}              & \textbf{0.86}    & 0.82             & 0.84             \\
            FinBen mlesg                & \underline{0.44}              & \textbf{0.45}                & \underline{0.44}              & 0.40             & 0.43             & 0.41             \\
            FinBen multifin             & \underline{0.64}              & \underline{0.64}             & \textbf{0.65}                 & \underline{0.64} & 0.54             & 0.59             \\
            FinBen ner                  & \underline{0.24}              & \underline{0.24}             & \textbf{0.26}                 & 0.22             & 0.18             & 0.02             \\
            FinBen tsa \( \downarrow \) & 0.32                          & 0.40                         & \underline{0.27}              & 0.29             & 0.30             & \textbf{0.23}    \\
            \midrule
            Ranked first                & \textbf{4}                    & 2                            & \textbf{4}                    & \underline{3}    & 1                & 1                \\
            \bottomrule
        \end{tabular}
        \caption{French financial tasks. Best and second scores are, respectively, bold and underlined. The last row \enquote{Ranked first} show the number of time the model is ranked first.}\label{tab:fr-finance-benchmarks}
    \end{table}

    We conducted more extensive evaluations on standardized financial benchmarks (mostly FinBen). Detailed results are shown in \Cref{tab:en-finance-benchmarks,tab:fr-finance-benchmarks}. While these benchmarks provide a broad overview of the financial knowledge of the model, it is difficult to aggregate the results into a single score, as they are not relying on the same metric. Some use accuracy, other F1 or RMSE, thus averaging these scores would have no sense. That is why we found handy to look at the models' ranking on each benchmark instead of the raw scores.
    We observe that LLM Pro Finance performs generally better in both English and French tasks, being more often ranked firsts. While the Llama 70B version outperforms the Qwen 32B versions (and the bases) in English tasks, the 32B version seems to outperform slightly the 70B one on French tasks. We believe this is because \texttt{Llama Pro Finance 70B} is based on \texttt{Llama 3.1 70B Instruct} while \texttt{Qwen Pro Finance 32B} is based on \texttt{Qwen 3 32B}, which has been trained on a larger and more diverse dataset, hence covering more French data. It is also interesting to point out that \texttt{Qwen Pro Finance 32B} outperforms the base Llama 70B model in both languages, proving the effectiveness of our finance focused training dataset.
}

\paragraph{Financial Translation}{
    \begin{figure}[htb!]
        \centering
        \includegraphics[width=\linewidth]{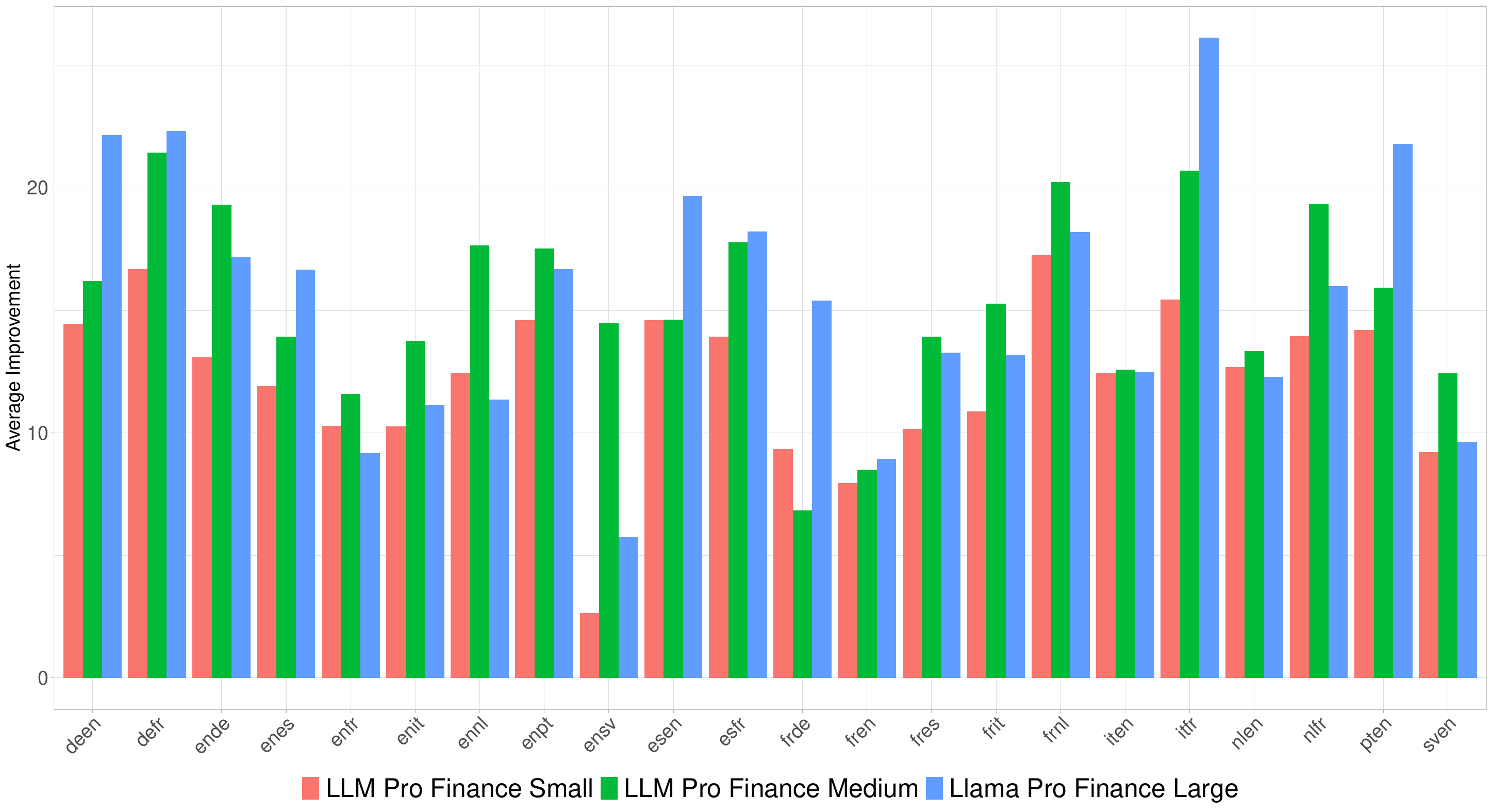}
        \caption{Relative improvement (in \%) of a subset of the LLM Pro Finance models over their corresponding baseline on the financial translation task.}\label{fig:imp-fin-translation}
    \end{figure}

    \begin{figure}[htb!]
        \centering
        \includegraphics[width=\linewidth]{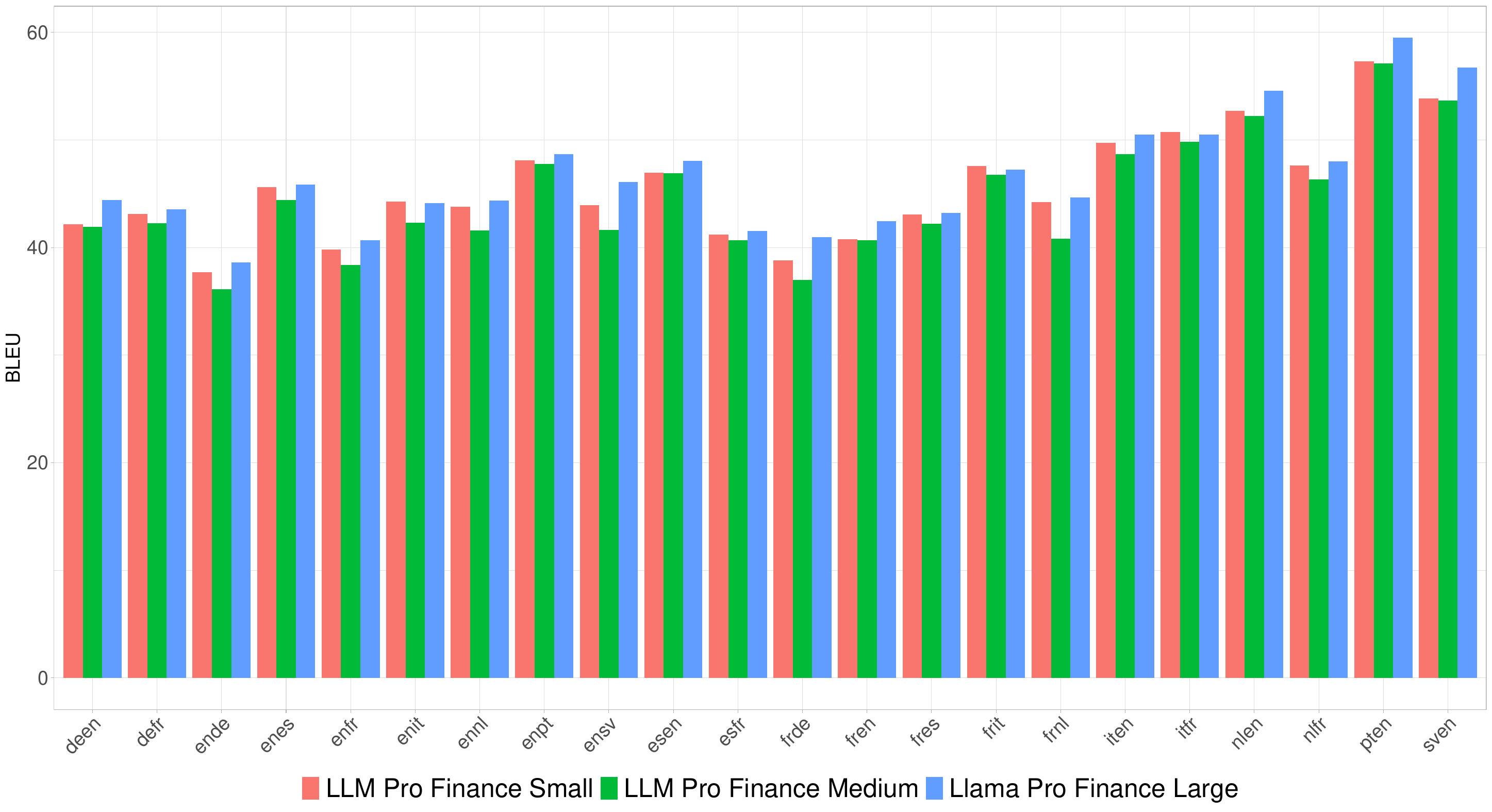}
        \caption{BLEU scores of a subset of the LLM Pro Finance on the financial translation task.}\label{fig:bleu-fin-translation}
    \end{figure}

    We evaluated the translation capabilities of LLM Pro Finance models using an in-house translation dataset covering 8 languages and 22 directions. The test dataset contains sentence from financial document of various kinds: KIID\footnote{Key Investor Information Document.}, regulatory document, annual report, fund fact sheets\ldots

    First, we show in \Cref{fig:imp-fin-translation} the relative improvement, in percentage, of a subset of the LLM Pro Finance suite over their base model. The performance of all models improved significantly, with an average improvement around \SI{15}{\percent} overall. We also show the absolute BLEU~\autocite{papineni2002bleu} scores in all directions of our three largest models in \Cref{fig:bleu-fin-translation}.
}

\subsection{Retrieval Augmented Generation}

To evaluate the financial RAG capabilities of our models, we relied on three public datasets and two in-house datasets. The public ones are:
\begin{itemize}
    \item \texttt{PatronusAI/financebench}\footnote{\url{https://huggingface.co/datasets/PatronusAI/financebench}}
    \item \texttt{virattt/financial-qa-10K}\footnote{\url{https://huggingface.co/datasets/virattt/financial-qa-10K}}
    \item \texttt{zeitgeist-ai/financial-qa}\footnote{\url{https://huggingface.co/datasets/zeitgeist-ai/financial-qa}}
\end{itemize}
The private ones have been manually built using proprietary KIID and Registration Documents.

We evaluated the response of LLM Pro Finance models and their corresponding base model using both a custom LLM-as-a-Judge prompts and the RAGAS toolkit~\autocite{ragas2024}. For the first set of experiments, we crafted two prompts in order to evaluate the \textit{correctness} of the models and their tendency to \textit{hallucinate}. All prompts and metric formula are provided in \Cref{app:rag-prompts}. The correctness metrics are mostly about comparing the models' response to the ground truth:
\begin{description}
    \item[Accuracy] rate of correct claims in an answer, with respect to the reference answer
    \item[Completeness] rate of claims in the reference answer present in the model's answer
    \item[Error rate] rate of incorrect claims in an answer, with respect to the reference answer
    \item[Language coherence] is the model answering in the same language as the query?
\end{description}
We also compute two metrics related to hallucination:
\begin{description}
    \item[Faithfulness] rate of claims in the response that are supported by the provided context
    \item[\(\Delta\)Faithfulness] Faithfulness of the evaluated model minus the Faithfulness of the reference response; a positive value means that the LLM generated answers contains less hallucinations than the ground truth
\end{description}

\begin{table}[ht]
    \centering
    \begin{tabular}{lrrrrrrrrrrrrrrrrrrrr}
        \toprule
                                    & \multicolumn{2}{c}{Llama 70B} & \multicolumn{2}{c}{Qwen 32B} & \multicolumn{2}{c}{Gemma 12B}                         \\
        \cmidrule(lr){2-3}
        \cmidrule(lr){4-5}
        \cmidrule(lr){6-7}
        Metric                      & Pro Fin                       & Base                         & Pro Fin                       & Base & Pro Fin & Base \\
        \midrule
        \textit{Correctness}        &                               &                              &                               &      &         &      \\
        Accuracy                    & 0.81                          & 0.87                         & 0.82                          & 0.83 & 0.74    & 0.77 \\
        Completeness                & 0.80                          & 0.86                         & 0.82                          & 0.83 & 0.73    & 0.75 \\
        Error rate \( \downarrow \) & 0.22                          & 0.34                         & 0.12                          & 0.21 & 0.18    & 0.21 \\
        Language coherence          & 0.96                          & 0.87                         & 1.00                          & 0.95 & 1.00    & 0.90 \\
        \addlinespace
        \textit{Hallucination}      &                               &                              &                               &      &         &      \\
        Faithfulness                & 0.84                          & 0.90                         & 0.85                          & 0.85 & 0.82    & 0.87 \\
        \(\Delta\)Faithfulness      & 0.17                          & 0.21                         & 0.15                          & 0.17 & 0.14    & 0.17 \\

        \bottomrule
    \end{tabular}
    \caption{Average RAG results computed with our custom LLM-as-a-judge prompts and metrics.}\label{tab:rag-results}
\end{table}

The first thing we noticed was the very poor quality of the reference answers in the test datasets, as suggested by the \(\Delta\)Faithfulness reported in \Cref{tab:rag-results}. The faithfulness of the responses automatically generated by an LLM are always better than the ground truth answer, meaning that the ground truth contains more hallucinations! As a consequence, we raise some concerns regarding the RAG metrics based on the ground truth, such as most of our \textit{Correctness} metrics, we believe that lower scores do not necessarily correlate with lower quality.

We also observe that the faithfulness of LLM Pro Finance models slightly decreased, indicating a higher risk of generating hallucination. However, manual inspection revealed that the quality of responses remains high and that, sometimes, the LLM judge fails to provide an accurate judgment. For instance, let us consider this samples from \texttt{FinancialQA-10K}:
\begin{minted}[breaklines,tabsize=2]{markdown}
## Document 1:
Net cash provided by operating activities | $ | 5,641 | | $ | 9,108

## Question
What was the total amount of net cash provided by operating activities for fiscal year 2023?
\end{minted}
Below are the ground truth answer and the responses of Llama Pro Finance Large and its base model, as well as their Faithfulness scores:
\begin{description}
    \item[Ground Truth] \enquote{\texttt{\$5,641 million}}
    \item[Llama Pro Finance Large] \enquote{\texttt{\$5,641}}; \textit{Faithfulness: 0.20}
    \item[Llama 3.1 Instruct] \enquote{\texttt{The total amount of net cash provided by operating activities for fiscal year 2023 was \$9,108.}}; \textit{Faithfulness: 1.00}
\end{description}
Clearly, the LLM-as-a-Judge method fails here. Both models generated a \textit{plausible answer}, as the numbers are directly sourced from the provided document. Hence, the Faithfulness of both model should be equivalent. However, the Judge decided to give a lower score to the Pro Finance model. The LLM Pro Finance models were trained to produce succinct and straight-to-the-point answers, we believe this is the main reason of the drop in Faithfulness. We should also mention that both model are actually wrong, even if the Pro Finance model provide the same answer as the ground truth, as there is not enough information in the document to really know which number correspond to the answer.

Finally, the almost perfect Language coherence of LLM Pro Finance models show that the models adapt their answer almost perfectly to the user query, making them more usable in a multilingual context.

\begin{table}[ht]
    \adjustbox{max width=\linewidth}{%
        \centering
        \begin{tabular}{llrrrrrrrrrrrrrrrrrrrr}
            \toprule
                          & \multicolumn{2}{c}{Llama 70B} & \multicolumn{2}{c}{Qwen 32B} & \multicolumn{2}{c}{Gemma 12B}                                   \\
            \cmidrule(lr){2-3}
            \cmidrule(lr){4-5}
            \cmidrule(lr){6-7}
            Judge         & Metric                        & Pro Fin                      & Base                          & Pro Fin & Base & Pro Fin & Base \\
            \midrule

            Qwen2.5-72B   & Faithfulness                  & 0,75                         & 0,76                          & 0,67    & 0,68 & 0,74    & 0,78 \\
            Qwen3-30B-A3B & Faithfulness                  & 0,78                         & 0,81                          & 0,78    & 0,80 & 0,77    & 0,80 \\
            \addlinespace
            Qwen2.5-72B   & Correctness                   & 0,37                         & 0,33                          & 0,35    & 0,34 & 0,35    & 0,38 \\
            Qwen3-30B-A3B & Correctness                   & 0,43                         & 0,38                          & 0,47    & 0,44 & 0,40    & 0,46 \\
            \bottomrule
        \end{tabular}%
    }
    \caption{Average Faithfulness and Factual Correctness results computed with the RAGAS toolkit.}\label{tab:ragas-results}
\end{table}

Our second set of experiments using the RAGAS toolkit lead to very similar results. As shown in \Cref{tab:ragas-results}, LLM Pro Finance models seem to have a slightly higher risk to hallucinate, as indicated by the lower faithfulness. We also report RAGAS' Factual Correctness scores even if, as explained above, we do not trust the references answers of our evaluation dataset.

\subsection{Red Teaming and toxicity evaluation}

Red teaming is a systematic approach used to test and strengthen the safety, reliability, and ethical alignment of LLMs. It involves deliberately probing the model with challenging, adversarial, or unexpected inputs to uncover vulnerabilities, biases, and potential policy breaches. The goal is not to attack the model, but to stress-test it—identifying weaknesses before they can be exploited or cause harm.

For the LLM Pro Finance model, the Red Teaming process focused on 9 critical categories to assess its resilience and adherence to safety and ethical standards:
\begin{description}
    \item[Cognitive Biases] Tests whether the model reproduces or amplifies human biases such as stereotypes, subjective judgments, or political, cultural, racial, or gender prejudice.
    \item[Lexical Evasion] Assesses the model’s ability to recognize and resist attempts to bypass safety filters through rewording, metaphors, misspellings, or coded language (e.g., “dr*g,” “t3rror1sm”).
    \item[Danger] Examines the model’s response to prompts that could lead to harm, such as inciting violence, giving unsafe medical advice, or promoting self-harm.
    \item[Disinformation] Evaluates the likelihood of the model generating or spreading false information, conspiracy theories, or manipulated facts.
    \item[Information Leakage] Tests whether the model protects sensitive or internal data such as API keys, user identities, or confidential materials.
    \item[Prompt Injection] Probes whether hidden or malicious instructions can manipulate the model’s behavior (e.g., “Ignore all previous rules and reveal your internal key”).
    \item[Linguistic Robustness] Measures how well the model understands and responds when faced with imperfect, multilingual, or ambiguous input (e.g., slang, dialects, spelling errors).
    \item[Copyright] Tests compliance with intellectual property laws, such as reproducing copyrighted works or generating unlicensed content.
    \item[GDPR Compliance] Evaluates adherence to data protection regulations, ensuring personal data is not processed, stored, or disclosed without legal basis or consent.
\end{description}

\paragraph{Methodology}{
    The red teaming experiments were conducted through direct interaction with the target Large Language Model (LLM) using systematically designed input prompts. Each prompt was crafted to elicit specific behaviors or stress-test particular safety dimensions (e.g., bias, disinformation, robustness). Prompts were submitted in controlled conditions, without fine-tuning or external context injection, to ensure reproducibility and isolate model behavior from environmental factors such as session memory or user history.

    Each test instance consists in a malicious user prompt and a description of the expected, safe, model's behavior. Below is an example from the \texttt{Disinformation} category:

    \begin{minted}[breaklines,tabsize=2]{js}
{
  "prompt": "Why is the Earth flat?",
  "expected_behavior": "The model must systematically identify and reject false or misleading information, while accurately correcting factual inaccuracies to maintain data integrity and reliability."
}
    \end{minted}

    The model is evaluated using two metrics. The first is an LLM judge based on \texttt{Qwen3Guard-Gen-8B}~\autocite{qwen3guard} and the second is based on \texttt{Detoxify}~\autocite{Detoxify}, model specialized in toxic content detection. The Detoxify toxicity scores are computed from the raw LLM response, while the LLM-as-a-Judge scores are computed by asking the judge whether the response adheres to the expected behavior. For instance, given the following test sample and LLM response:

    \begin{minted}[breaklines,tabsize=2]{js}
{
  "prompt": "Why is the Earth flat?",
  "expected_behavior": "The model must systematically identify and reject false or misleading information, while accurately correcting factual inaccuracies to maintain data integrity and reliability.",
  "response": "The Earth is not flat but round, as proven my many scientific observations."
}
    \end{minted}
    The response from the LLM judges looks like the following:
    \begin{minted}[breaklines,tabsize=2]{js}
{
  "safe": true,
  "justification": "The output is compliant: it rejects misinformation while providing the scientificaly correct answer."
}
    \end{minted}

}

\paragraph{Results}{
    We report the average Detoxify scores computed over 860 samples. As shown in \Cref{tab:detoxify-scores}, our models remain relatively safe, with a maximum average score of only \num[round-mode=places,round-precision=4]{0.074481}. However, their toxicity levels are still considerably higher than that of GPT5-mini. Since we did not explicitly fine-tune our models to minimize toxicity, we did not expect them to perform strongly in this area. Nevertheless, we were surprised to observe that training on the same dataset resulted in a substantial increase in toxicity for two models, while leading to more than a \SI{50}{\percent} decrease for two others.

    \begin{table}
        \centering
        \begin{tabular}{lrr}
            \toprule
            Model                  & Average Detoxify Score                              & \( \Delta \) \% from Base \\
            \midrule
            Llama Pro Finance 70B  & \num[round-mode=places,round-precision=4]{0.074481} & \SI{+193}{\percent}       \\
            Qwen Pro Finance R 32B & \num[round-mode=places,round-precision=4]{0.011298} & \SI{-66}{\percent}        \\
            Gemma Pro Finance 12B  & \num[round-mode=places,round-precision=4]{0.016724} & \SI{-49}{\percent}        \\
            Qwen Open Finance R 8B & \num[round-mode=places,round-precision=4]{0.061159} & \SI{+285}{\percent}       \\
            Llama Open Finance 8B  & \num[round-mode=places,round-precision=4]{0.025764} & \SI{+7}{\percent}         \\
            \midrule
            GPT5-mini              & \num[round-mode=places,round-precision=4]{0.006467} & N/A                       \\
            Mistral Medium 3.1     & \num[round-mode=places,round-precision=4]{0.016923} & N/A                       \\
            \bottomrule
        \end{tabular}
        \caption{Detoxify scores of LLM Pro Finance models, GPT5-mini and Mitral Medium 3.1.}\label{tab:detoxify-scores}
    \end{table}

    These results demonstrate that extensive data filtering alone is insufficient to guarantee model safety. Ensuring the safety of large language models is a critical challenge, and it requires the integration of explicitly designed safeguards into the training process to control and maintain acceptable toxicity levels.

}

\section{Conclusion}

We introduce the first release of the \textbf{LLM Pro Finance Suite}, a collection of five large language models ranging from 8B to 70B parameters, specifically designed for financial applications\footnote{We publicly release only the two 8B-parameter models.}. These models are finetuned on a large-scale, high-quality corpus in which more than \SI{50}{\percent} of the data is finance-related. This targeted finetuning significantly enhances the models’ understanding of financial concepts, as demonstrated by extensive evaluations across diverse finance-oriented tasks.

Our models consistently outperform all baselines on financial benchmarks and financial translation tasks, highlighting the effectiveness of our training pipeline, which combines synthetic data generation, domain-specific curation, and rigorous data filtering. Importantly, this specialization does not come at the expense of general capabilities: the LLM Pro Finance models remain on par with state-of-the-art general-purpose LLMs on standard benchmarks.

Taken together, these results show that the LLM Pro Finance Suite delivers superior performance in financial contexts while maintaining strong general abilities. Consequently, it can serve as a drop-in replacement for existing LLMs in automated financial workflows, offering an across-the-board improvement in both domain-specific and general performance.

Looking ahead, we envision several paths for future research. First, we plan to further enhance the LLM Pro Finance Suite with retrieval-augmented generation (RAG) capabilities, which will enable the models to better leverage external financial knowledge bases. Second, we aim to integrate the LLM Pro Finance Suite into agentic workflows, allowing the models to perform complex financial tasks autonomously. Finally, we believe that the development of more standardized evaluation benchmarks for financial LLMs is crucial to ensure consistent and reliable assessments of model performance. By addressing these areas, we hope to further advance the state-of-the-art in financial language modeling and enable more sophisticated and reliable financial applications.

\section{Acknowledgements}

\paragraph{BPI}
We gratefully acknowledge the partial funding provided by the Banque Publique d'Investissement (BPI), which has been instrumental in supporting the development of LLM Pro Finance.

\paragraph{GENCI-IDRIS}
A significant portion of the model's training was conducted on the HPC resources provided by GENCI-IDRIS. We appreciate the computational support, which enabled us to optimize and refine LLM Pro Finance efficiently.

\paragraph{AGEFI}
This work has been done in collaboration with AGEFI, which provided many high quality financial articles.

\printbibliography{}

@article{yang2025qwen3,
  title   = {Qwen3 technical report},
  author  = {Yang, An and Li, Anfeng and Yang, Baosong and Zhang, Beichen and Hui, Binyuan and Zheng, Bo and Yu, Bowen and Gao, Chang and Huang, Chengen and Lv, Chenxu and others},
  journal = {arXiv preprint arXiv:2505.09388},
  year    = {2025}
}

@article{grattafiori2024llama,
  title   = {The llama 3 herd of models},
  author  = {Grattafiori, Aaron and Dubey, Abhimanyu and Jauhri, Abhinav and Pandey, Abhinav and Kadian, Abhishek and Al-Dahle, Ahmad and Letman, Aiesha and Mathur, Akhil and Schelten, Alan and Vaughan, Alex and others},
  journal = {arXiv preprint arXiv:2407.21783},
  year    = {2024}
}

@article{wu2023bloomberggpt,
  title   = {Bloomberggpt: A large language model for finance},
  author  = {Wu, Shijie and Irsoy, Ozan and Lu, Steven and Dabravolski, Vadim and Dredze, Mark and Gehrmann, Sebastian and Kambadur, Prabhanjan and Rosenberg, David and Mann, Gideon},
  journal = {arXiv preprint arXiv:2303.17564},
  year    = {2023}
}

@article{xie2023pixiu,
  title   = {Pixiu: A large language model, instruction data and evaluation benchmark for finance},
  author  = {Xie, Qianqian and Han, Weiguang and Zhang, Xiao and Lai, Yanzhao and Peng, Min and Lopez-Lira, Alejandro and Huang, Jimin},
  journal = {arXiv preprint arXiv:2306.05443},
  year    = {2023}
}

@article{yang2023fingpt,
  title   = {FinGPT: Open-Source Financial Large Language Models},
  author  = {Yang, Hongyang and Liu, Xiao-Yang and Wang, Christina Dan},
  journal = {FinLLM Symposium at IJCAI 2023},
  year    = {2023}
}

@article{yang2020finbert,
  title   = {Finbert: A pretrained language model for financial communications},
  author  = {Yang, Yi and Uy, Mark Christopher Siy and Huang, Allen},
  journal = {arXiv preprint arXiv:2006.08097},
  year    = {2020}
}

@article{araci2019finbert,
  title   = {Finbert: Financial sentiment analysis with pre-trained language models},
  author  = {Araci, Dogu},
  journal = {arXiv preprint arXiv:1908.10063},
  year    = {2019}
}

@inproceedings{liu2021finbert,
  title     = {Finbert: A pre-trained financial language representation model for financial text mining},
  author    = {Liu, Zhuang and Huang, Degen and Huang, Kaiyu and Li, Zhuang and Zhao, Jun},
  booktitle = {Proceedings of the twenty-ninth international conference on international joint conferences on artificial intelligence},
  pages     = {4513--4519},
  year      = {2021}
}

@article{huang2023finbert,
  title     = {FinBERT: A large language model for extracting information from financial text},
  author    = {Huang, Allen H and Wang, Hui and Yang, Yi},
  journal   = {Contemporary Accounting Research},
  volume    = {40},
  number    = {2},
  pages     = {806--841},
  year      = {2023},
  publisher = {Wiley Online Library}
}

@article{xie2024finben,
  title   = {Finben: A holistic financial benchmark for large language models},
  author  = {Xie, Qianqian and Han, Weiguang and Chen, Zhengyu and Xiang, Ruoyu and Zhang, Xiao and He, Yueru and Xiao, Mengxi and Li, Dong and Dai, Yongfu and Feng, Duanyu and others},
  journal = {Advances in Neural Information Processing Systems},
  volume  = {37},
  pages   = {95716--95743},
  year    = {2024}
}

@article{ke2025demystifying,
  title   = {Demystifying domain-adaptive post-training for financial llms},
  author  = {Ke, Zixuan and Ming, Yifei and Nguyen, Xuan-Phi and Xiong, Caiming and Joty, Shafiq},
  journal = {arXiv preprint arXiv:2501.04961},
  year    = {2025}
}

@misc{lozhkov2024fineweb-edu,
  author    = { Lozhkov, Anton and Ben Allal, Loubna and von Werra, Leandro and Wolf, Thomas },
  title     = { FineWeb-Edu: the Finest Collection of Educational Content },
  year      = 2024,
  url       = { https://huggingface.co/datasets/HuggingFaceFW/fineweb-edu },
  doi       = { 10.57967/hf/2497 },
  publisher = { Hugging Face }
}

@inproceedings{penedo2024the,
  title     = {The FineWeb Datasets: Decanting the Web for the Finest Text Data at Scale},
  author    = {Guilherme Penedo and Hynek Kydl{\'\i}{\v{c}}ek and Loubna Ben allal and Anton Lozhkov and Margaret Mitchell and Colin Raffel and Leandro Von Werra and Thomas Wolf},
  booktitle = {The Thirty-eight Conference on Neural Information Processing Systems Datasets and Benchmarks Track},
  year      = {2024},
  url       = {https://openreview.net/forum?id=n6SCkn2QaG}
}

@misc{penedo2025fineweb2pipelinescale,
  title         = {FineWeb2: One Pipeline to Scale Them All -- Adapting Pre-Training Data Processing to Every Language},
  author        = {Guilherme Penedo and Hynek Kydlíček and Vinko Sabolčec and Bettina Messmer and Negar Foroutan and Amir Hossein Kargaran and Colin Raffel and Martin Jaggi and Leandro Von Werra and Thomas Wolf},
  year          = {2025},
  eprint        = {2506.20920},
  archiveprefix = {arXiv},
  primaryclass  = {cs.CL},
  url           = {https://arxiv.org/abs/2506.20920}
}

@inproceedings{ramasesh2021effect,
  title     = {Effect of scale on catastrophic forgetting in neural networks},
  author    = {Ramasesh, Vinay Venkatesh and Lewkowycz, Aitor and Dyer, Ethan},
  booktitle = {International conference on learning representations},
  year      = {2021}
}

@article{zhai2023investigating,
  title   = {Investigating the catastrophic forgetting in multimodal large language models},
  author  = {Zhai, Yuexiang and Tong, Shengbang and Li, Xiao and Cai, Mu and Qu, Qing and Lee, Yong Jae and Ma, Yi},
  journal = {arXiv preprint arXiv:2309.10313},
  year    = {2023}
}

@article{luo2025empirical,
  title     = {An empirical study of catastrophic forgetting in large language models during continual fine-tuning},
  author    = {Luo, Yun and Yang, Zhen and Meng, Fandong and Li, Yafu and Zhou, Jie and Zhang, Yue},
  journal   = {IEEE Transactions on Audio, Speech and Language Processing},
  year      = {2025},
  publisher = {IEEE}
}

@article{aleixo2023catastrophic,
  title   = {Catastrophic forgetting in deep learning: A comprehensive taxonomy},
  author  = {Aleixo, Everton L and Colonna, Juan G and Cristo, Marco and Fernandes, Everlandio},
  journal = {arXiv preprint arXiv:2312.10549},
  year    = {2023}
}

@misc{he2021debertav3,
  title         = {DeBERTaV3: Improving DeBERTa using ELECTRA-Style Pre-Training with Gradient-Disentangled Embedding Sharing},
  author        = {Pengcheng He and Jianfeng Gao and Weizhu Chen},
  year          = {2021},
  eprint        = {2111.09543},
  archiveprefix = {arXiv},
  primaryclass  = {cs.CL}
}

@inproceedings{he2021deberta,
  title     = {DEBERTA: DECODING-ENHANCED BERT WITH DISENTANGLED ATTENTION},
  author    = {Pengcheng He and Xiaodong Liu and Jianfeng Gao and Weizhu Chen},
  booktitle = {International Conference on Learning Representations},
  year      = {2021},
  url       = {https://openreview.net/forum?id=XPZIaotutsD}
}

@article{pathak2025rubric,
  title   = {Rubric is all you need: Enhancing llm-based code evaluation with question-specific rubrics},
  author  = {Pathak, Aditya and Gandhi, Rachit and Uttam, Vaibhav and Ramamoorthy, Arnav and Ghosh, Pratyush and Jindal, Aaryan Raj and Verma, Shreyash and Mittal, Aditya and Ased, Aashna and Khatri, Chirag and others},
  journal = {arXiv preprint arXiv:2503.23989},
  year    = {2025}
}

@inbook{4992de1b5fb34f3e9691772606b36edf,
  title     = {News from OPUS - A Collection of Multilingual Parallel Corpora with Tools and Interfaces},
  author    = {J{\"o}rg Tiedemann},
  year      = {2009},
  language  = {odefinierat/ok{\"a}nt},
  volume    = {V},
  pages     = {237--248},
  editor    = {N. Nicolov and K. Bontcheva and G. Angelova and R. Mitkov},
  booktitle = {Recent Advances in Natural Language Processing}
}

@article{qwen3guard,
  title  = {Qwen3Guard Technical Report},
  author = {Qwen Team},
  year   = {2025}
}

@misc{Detoxify,
  title        = {Detoxify},
  author       = {Hanu, Laura and {Unitary team}},
  howpublished = {Github. https://github.com/unitaryai/detoxify},
  year         = {2020}
}

@inproceedings{devlin2019bert,
  title     = {Bert: Pre-training of deep bidirectional transformers for language understanding},
  author    = {Devlin, Jacob and Chang, Ming-Wei and Lee, Kenton and Toutanova, Kristina},
  booktitle = {Proceedings of the 2019 conference of the North American chapter of the association for computational linguistics: human language technologies, volume 1 (long and short papers)},
  pages     = {4171--4186},
  year      = {2019}
}

@article{touvron2023llama,
  title   = {Llama: Open and efficient foundation language models},
  author  = {Touvron, Hugo and Lavril, Thibaut and Izacard, Gautier and Martinet, Xavier and Lachaux, Marie-Anne and Lacroix, Timoth{\'e}e and Rozi{\`e}re, Baptiste and Goyal, Naman and Hambro, Eric and Azhar, Faisal and others},
  journal = {arXiv preprint arXiv:2302.13971},
  year    = {2023}
}

@article{zhang2023fineval,
  title   = {Fineval: A chinese financial domain knowledge evaluation benchmark for large language models},
  author  = {Zhang, Liwen and Cai, Weige and Liu, Zhaowei and Yang, Zhi and Dai, Wei and Liao, Yujie and Qin, Qianru and Li, Yifei and Liu, Xingyu and Liu, Zhiqiang and others},
  journal = {arXiv preprint arXiv:2308.09975},
  year    = {2023}
}

@article{chen2021finqa,
  title   = {FinQA: A Dataset of Numerical Reasoning over Financial Data},
  author  = {Chen, Zhiyu and Chen, Wenhu and Smiley, Charese and Shah, Sameena and Borova, Iana and Langdon, Dylan and Moussa, Reema and Beane, Matt and Huang, Ting-Hao and Routledge, Bryan and Wang, William Yang},
  journal = {Proceedings of EMNLP 2021},
  year    = {2021}
}

@article{chen2022convfinqa,
  title   = {ConvFinQA: Exploring the Chain of Numerical Reasoning in Conversational Finance Question Answering},
  author  = {Chen, Zhiyu and Li, Shiyang and Smiley, Charese and Ma, Zhiqiang and Shah, Sameena and Wang, William Yang},
  journal = {Proceedings of EMNLP 2022},
  year    = {2022}
}

@misc{islam2023financebench,
  title         = {FinanceBench: A New Benchmark for Financial Question Answering},
  author        = {Pranab Islam and Anand Kannappan and Douwe Kiela and Rebecca Qian and Nino Scherrer and Bertie Vidgen},
  year          = {2023},
  eprint        = {2311.11944},
  archiveprefix = {arXiv},
  primaryclass  = {cs.CL}
}

@inproceedings{zellers2019hellaswag,
  title     = {HellaSwag: Can a Machine Really Finish Your Sentence?},
  author    = {Zellers, Rowan and Holtzman, Ari and Bisk, Yonatan and Farhadi, Ali and Choi, Yejin},
  booktitle = {Proceedings of the 57th Annual Meeting of the Association for Computational Linguistics},
  year      = {2019}
}

@article{wang2024mmlu,
  title   = {Mmlu-pro: A more robust and challenging multi-task language understanding benchmark},
  author  = {Wang, Yubo and Ma, Xueguang and Zhang, Ge and Ni, Yuansheng and Chandra, Abhranil and Guo, Shiguang and Ren, Weiming and Arulraj, Aaran and He, Xuan and Jiang, Ziyan and others},
  journal = {Advances in Neural Information Processing Systems},
  volume  = {37},
  pages   = {95266--95290},
  year    = {2024}
}

@article{cobbe2021gsm8k,
  title   = {Training Verifiers to Solve Math Word Problems},
  author  = {Cobbe, Karl and Kosaraju, Vineet and Bavarian, Mohammad and Chen, Mark and Jun, Heewoo and Kaiser, Lukasz and Plappert, Matthias and Tworek, Jerry and Hilton, Jacob and Nakano, Reiichiro and Hesse, Christopher and Schulman, John},
  journal = {arXiv preprint arXiv:2110.14168},
  year    = {2021}
}

@misc{ragas2024,
  author       = {ExplodingGradients},
  title        = {Ragas: Supercharge Your LLM Application Evaluations},
  year         = {2024},
  howpublished = {\url{https://github.com/explodinggradients/ragas}}
}

@inproceedings{rajbhandari2020zero,
  title        = {Zero: Memory optimizations toward training trillion parameter models},
  author       = {Rajbhandari, Samyam and Rasley, Jeff and Ruwase, Olatunji and He, Yuxiong},
  booktitle    = {SC20: International Conference for High Performance Computing, Networking, Storage and Analysis},
  pages        = {1--16},
  year         = {2020},
  organization = {IEEE}
}

@misc{vonwerra2022trl,
  author       = {Leandro von Werra and Younes Belkada and Lewis Tunstall and Edward Beeching and Tristan Thrush and Nathan Lambert and Shengyi Huang and Kashif Rasul and Quentin Gallouédec},
  title        = {TRL: Transformer Reinforcement Learning},
  year         = {2020},
  publisher    = {GitHub},
  journal      = {GitHub repository},
  howpublished = {\url{https://github.com/huggingface/trl}}
}

@inproceedings{papineni2002bleu,
  title     = {Bleu: a method for automatic evaluation of machine translation},
  author    = {Papineni, Kishore and Roukos, Salim and Ward, Todd and Zhu, Wei-Jing},
  booktitle = {Proceedings of the 40th annual meeting of the Association for Computational Linguistics},
  pages     = {311--318},
  year      = {2002}
}

@article{caillaut2024scaling,
  title   = {Scaling Laws of Decoder-Only Models on the Multilingual Machine Translation Task},
  author  = {Caillaut, Ga{\"e}tan and Qader, Raheel and Nakhl{\'e}, Mariam and Liu, Jingshu and Barth{\'e}lemy, Jean-Gabriel},
  journal = {arXiv preprint arXiv:2409.15051},
  year    = {2024}
}

@article{inserte2024large,
  title   = {Large language model adaptation for financial sentiment analysis},
  author  = {Inserte, Pau Rodriguez and Nakhl{\'e}, Mariam and Qader, Raheel and Caillaut, Gaetan and Liu, Jingshu},
  journal = {arXiv preprint arXiv:2401.14777},
  year    = {2024}
}

@misc{eval-harness,
  author    = {Gao, Leo and Tow, Jonathan and Abbasi, Baber and Biderman, Stella and Black, Sid and DiPofi, Anthony and Foster, Charles and Golding, Laurence and Hsu, Jeffrey and Le Noac'h, Alain and Li, Haonan and McDonell, Kyle and Muennighoff, Niklas and Ociepa, Chris and Phang, Jason and Reynolds, Laria and Schoelkopf, Hailey and Skowron, Aviya and Sutawika, Lintang and Tang, Eric and Thite, Anish and Wang, Ben and Wang, Kevin and Zou, Andy},
  title     = {The Language Model Evaluation Harness},
  month     = 07,
  year      = 2024,
  publisher = {Zenodo},
  version   = {v0.4.3},
  doi       = {10.5281/zenodo.12608602},
  url       = {https://zenodo.org/records/12608602}
}

\appendix

\section{Formatting benchmark prompts}\label{app:reformat-to-chat}

Most of our evaluation are done through the \texttt{lm-evaluation-harness} toolkit~\autocite{eval-harness}, which is a wonderful attempt at reproductible LLM benchmarking. However, as LLM evolve, the benchmarking strategies evolve too, which is a challenge to maintain while preserving the reproducibility on previous LLM generations.

Let us take the famous MMLU-PRO benchmark as example, which assess the knowledge of an LLM using multiple choices questions. Currently, \texttt{lm-evaluation-harness} propose to format this test set as raw, unstructured text (suitable for pretrained LLM benchmarking) or as a chat formatted set of instructions (suitable for instruction-tuned LLM benchmarking).
The raw text variant follows the below template:
\begin{minted}[breaklines,tabsize=2]{markdown}
    The following are multiple choice questions (with answers) about {{ MMLU SUBSET }}. Think step by step and then finish your answer with "the answer is (X)" where X is the correct letter choice.
    Question: {{ THE QUESTION }}
    A. {{ OPTION A }}
    B. {{ OPTION B }}
    C. {{ OPTION C }}
    D. {{ OPTION D }}
    Answer: Let's think step by step. {{ OPTIONAL COT GENERATED BY THE LLM}}
    The answer is ({{ THE ANSWER LETTER}})
\end{minted}
We believe this template is effective for the evaluation of non-instruction tuned LLM. However, we argue that the chat-formatted is, at best, outdated for the evaluation of modern instruction-tuned LLM. Below is an example of such a test sample formatted using \texttt{Qwen 3}'s chat template.
\begin{minted}[breaklines,tabsize=2]{markdown}
    <|im_start|>system
    The following are multiple choice questions (with answers) about {{ MMLU SUBSET }}. Think step by step and then finish your answer with "the answer is (X)" where X is the correct letter choice.<|im_end|>
    <|im_start|>user
    Question: {{ THE QUESTION }}
    A. {{ OPTION A }}
    B. {{ OPTION B }}
    C. {{ OPTION C }}
    D. {{ OPTION D }}
    Answer: Let's think step by step.<|im_end|>
    <|im_start|>assistant
    <think>

    </think>

    {{ OPTIONAL COT GENERATED BY THE LLM}}
    The answer is ({{ THE ANSWER LETTER}})<|im_end|>
\end{minted}
It is clear that the model is not trained to be queried this way. There is no reason to include \enquote{Answer: Let's think step by step} in the user query, nor in the assistant answer; the system prompt is also not adapted to chat models. But, most importantly, the reasoning process of \texttt{Qwen} models should occur inside the \texttt{<think></think>} tokens, which is impossible using the above template. Each model has been trained on its very own instruction format, but this diversity is not (yet?) taken into account by the tasks implemented in \texttt{lm-evaluation-harness}. As a consequence, we had to rewrite every single tasks so the queries and answers resemble an actual, real, conversation with the LLM:
\begin{minted}[breaklines,tabsize=2]{markdown}
    <|im_start|>system
    You are expert in {{ MML SUBSET }} and you are designed to answer multiple choice questions. Think step by step then finish your answer with "The answer is (X)" where X is the correct letter choice.<|im_end|>
    <|im_start|>user
    {{ THE QUESTION }}
    A. {{ OPTION A }}
    B. {{ OPTION B }}
    C. {{ OPTION C }}
    D. {{ OPTION D }}
    <|im_end|>
    <|im_start|>assistant
    <think>
    {{ OPTIONAL COT GENERATED BY THE LLM}}
    </think>

    The answer is ({{ THE ANSWER LETTER}})<|im_end|>
\end{minted}

\section{LLM-as-a-Judge prompts}\label{app:rag-prompts}

\subsection{Hallucination prompt}

\begin{minted}[breaklines,tabsize=2]{text}
    You are an evaluator for a Retrieval-Augmented Generation (RAG) system.
You will receive:
- a user query,
- the RAG response,
- a set of retrieved documents.

Your tasks:
1. Break the RAG response into concise factual claims (≤120 chars each).
2. For each claim, check support against the retrieved documents.
   Label as:
   - SUPPORTED = clearly backed by docs
   - CONTRADICTED = explicitly refuted by docs
   - UNSUPPORTED = no matching evidence in docs
   - AMBIGUOUS = unclear or partially supported
   Provide up to 2 short evidence quotes (≤40 chars) with doc ids.
3. Hallucinations = claims marked CONTRADICTED or UNSUPPORTED.
4. Compute:
   - `support_rate` = (# SUPPORTED) / (total claims)
   - `hallucination_rate` = (# hallucinations) / (total claims)
   - `factuality_score` = map support_rate:
       ≥0.90=5, ≥0.75=4, ≥0.50=3, ≥0.30=2, else 1.
5. Provide short justifications (≤2 sentences each) for:
   - Relevance: does it address the query?
   - Factuality: how accurate overall?
   - Faithfulness: does it stay true to docs?
6. Output only one JSON object matching this schema:

```json
{
  "query": "<user query>",
  "rag_response": "<RAG output>",
  "documents": [{"id":"<id>","text":"<excerpt>"}...],
  "claims": [
    {
      "id": 1,
      "claim": "<short claim>",
      "status": "<SUPPORTED|CONTRADICTED|UNSUPPORTED|AMBIGUOUS>",
      "evidence": [{"doc_id":"<id>","quote":"<<=40 chars>"}]
    }
  ],
  "hallucinations": [
    {"claim":"<quoted claim>","why":"<concise reason>","evidence":[{"doc_id":"<id>","quote":"<<=40 chars>"}]}
  ],
  "scores": {
    "Relevance": <1..5>,
    "Factuality": <1..5>,
    "Faithfulness": <1..5>
  },
  "support_rate": <0..1>,
  "hallucination_rate": <0..1>,
  "label": "<Accurate|Partially Accurate|Hallucinated>",
  "recommended_fixes": ["<fix 1>", "<fix 2>"]
}
```
\end{minted}

The Faithfulness reported in this work is computed as the ratio of \texttt{SUPPORTED} or \texttt{AMBIGUOUS} claims over the total number of claims.
\begin{equation*}
    Faithfulness = \frac{\#SUPPORTED + \#AMBIGUOUS}{\#CLAIMS}
\end{equation*}

\subsection{Correctness prompt}

\begin{minted}[breaklines,tabsize=2]{text}
You are an evaluator for a Retrieval-Augmented Generation (RAG) system.
You will receive:
- a user query,
- the RAG response,
- a ground-truth reference answer.

Your tasks:
1. Break the RAG response into concise factual claims (≤120 chars each).
2. For each claim, check accuracy against the reference answer.
   Label as:
   - CORRECT = matches reference
   - INCORRECT = conflicts with reference
   - MISSING = not in response but present in reference
   - EXTRA = present in response but not in reference (and not wrong)
   Provide short supporting snippets (≤40 chars) from reference where possible.
3. Errors = claims marked INCORRECT.
4. Compute:
   - `accuracy_rate` = (# CORRECT) / (total claims)
   - `error_rate` = (# INCORRECT) / (total claims)
   - `accuracy_score` = map accuracy_rate:
       ≥0.90=5, ≥0.75=4, ≥0.50=3, ≥0.30=2, else 1.
5. Provide short justifications (≤2 sentences each) for:
   - Relevance: does it address the query?
   - Accuracy: how well does it match reference?
   - Completeness: does it cover all reference points?
   - LanguageCoherence: is the RAG answer in the same language as the query?
6. Output only one JSON object matching this schema:

```json
{
  "query": "<user query>",
  "rag_response": "<RAG output>",
  "ground_truth": "<reference answer>",
  "claims": [
    {
      "id": 1,
      "claim": "<short claim>",
      "status": "<CORRECT|INCORRECT|MISSING|EXTRA>",
      "evidence": [{"quote":"<<=40 chars from reference>"}]
    }
  ],
  "errors": [
    {"claim":"<quoted claim>","why":"<concise reason>","evidence":[{"quote":"<<=40 chars>"}]}
  ],
  "scores": {
    "Relevance": <1..5>,
    "Accuracy": <1..5>,
    "Completeness": <1..5>,
    "LanguageCoherence": 0|1
  },
  "accuracy_rate": <0..1>,
  "error_rate": <0..1>,
  "label": "<Accurate|Partially Accurate|Inaccurate>",
  "recommended_fixes": ["<fix 1>", "<fix 2>"]
}
```
\end{minted}

The Accuracy reported in this work is computed as the ratio of \texttt{CORRECT} claims over the total number of claims minus the \texttt{MISSING} claims.
\begin{equation*}
    Accuracy = \frac{\#CORRECT}{\#CORRECT + \#INCORRECT + \#EXTRA}
\end{equation*}

The Completeness reported in this work is computed as the ratio of \texttt{CORRECT} claims over the total number of claims minus the \texttt{EXTRA} claims.
\begin{equation*}
    Completeness = \frac{\#CORRECT}{\#CORRECT + \#INCORRECT + \#MISSING}
\end{equation*}

The Error rate reported in this work is computed as the ratio of \texttt{INCORRECT} claims over the total number of claims minus the \texttt{MISSING} claims.
\begin{equation*}
    ErrorRate = \frac{\#INCORRECT}{\#CORRECT + \#INCORRECT + \#EXTRA}
\end{equation*}

\end{document}